\documentclass[aps,prx,twocolumn,superscriptaddress,showpacs,floatfix,maxbibnames=10]{revtex4-2}
\usepackage{amsmath,amsthm,amssymb,amsfonts,dsfont,float,graphics,natbib,epsfig,epstopdf,color,verbatim,tabularx,bm,multirow,appendix}
\usepackage[margin=1.5cm]{geometry}
\usepackage[normalem]{ulem}

\usepackage{tikz}


\newcommand{\ket}[1]{| #1 \rangle}
\newcommand{\bra}[1]{\langle #1 |}
\newcommand{\tr}{\text{tr}}
\newcommand{\re}{\text{Re }}
\newcommand{\im}{\text{Im }}

\newcommand{\half}{\frac{1}{2}}
\newcommand{\quarter}{\frac{1}{4}}

\newcommand{\n}{\notag \\}
\newcommand{\ep}{\\}
\newcommand{\supp}{\cite{supp}~}

\begin{document}

\def\udem{D\'epartement de Physique, Universit\'e de Montr\'eal, Montr\'eal, QC, Canada H3C 3J7}
\def\udemB{Centre de Recherches Math\'ematiques, Universit\'e de Montr\'eal, Montr\'eal, QC, Canada H3C 3J7}
\def\udemC{Institut Courtois, Universit\'e de Montr\'eal, Montr\'eal, QC, Canada H2V 0B3}

\title{Fermionic Genuine Multiparty Entanglement}
\author{James Allen}
\email{james.allen@umontreal.ca}
\affiliation{\udem}
\author{Liuke Lyu}
\affiliation{\udem}
\author{William Witczak-Krempa}
\affiliation{\udem}
\affiliation{\udemB}
\affiliation{\udemC}
\date{\today}

\begin{abstract}
    Entanglement can show fundamentally different behavior in fermionic systems. However, while bipartite measures of fermionic entanglement have been established, genuine multiparty entanglement (GME) in fermionic systems is much less understood. We introduce an efficiently computable measure (via semi-definite programming) of fermionic GME, fermion genuine multiparty negativity (fGMN), and show that it is an entanglement monotone and a natural multipartite extension of the fermionic negativity. Using this measure, we find many states which have fGMN but no non-fermionic GME, including large classes of fermionic stabilizer states. The fGMN also displays some major phenomenological differences to the bipartite fermionic negativity, such as finite sudden death points over separation and temperature.
\end{abstract}

\maketitle

\section{Introduction} 

Entanglement is the defining characteristic of quantum systems. It is an essential resource for any quantum algorithm that aims to provide an advantage over classical counterparts, and it underpins the exotic behaviour of all strongly correlated quantum systems in condensed matter. Of particular interest is entanglement between more than two parties~\cite{Linden1998,Bennett1999,Seevinck2001,Jungnitsch2011}. Multipartite entanglement is an extremely diverse field where there are many types of entanglement that can arise~\cite{Dur2000,Acin2001,Verstraete2002}, one of the strongest being genuine multiparty entanglement (GME)~\cite{Toth2005}, in which all parties share in the entanglement together. GME represents entanglement in its most powerful, interesting form, and systems with GME have applications in measurement-based quantum computing~\cite{Raussendorf2001,Briegel2009}, quantum error correction~\cite{Scott2004}, quantum metrology~\cite{Sorensen2001,Giovannetti2004,Toth2012} and key distribution~\cite{Cabello2000,Epping2017}. However, it is also very difficult to characterize. Some entanglement measures, such as the geometric entanglement~\cite{Shimony1995,Barnum2001,Wei2003}, are faithful but difficult to compute, requiring the solution of an inefficient or intractable optimization problem to evaluate, while other measures such as those using specific witnesses~\cite{Horodecki1996,Horodecki2001,Guhne2010} can only guarantee entanglement on a very specific set of states.

More complications arise when we consider that many systems of interest, and in fact the vast majority of condensed matter systems, are described by fermionic degrees of freedom. The antisymmetric exchange statistics of fermions, and their parity superselection rule~\cite{Schliemann2001,Verstraete2003,Schuch2004,Banuls2007,Benatti2012,Benatti2014}, demand special considerations when determining if a particular fermionic system is entangled or not. Ref.~\onlinecite{Shapourian2017} successfully extended the negativity - an entanglement monotone for bipartite mixed states - to fermions, showing that fermionic entanglement is often more ubiquitous than non-fermionic measures. 
In particular, fermionic bipartite entanglement does not suffer a sudden death as a function of parameters such as temperature or separation. Interestingly, by studying the set of fermion-biseparable states it was found that fermionic GME does suffer a sudden death, thus sharing the same fate of entanglement as non-fermionic GME~\cite{Parez2026}. However, no measure of fermion GME was available to test and quantify the result.

In this paper, we devise the first efficiently computable measure of fermionic GME for mixed states, by modifying the renormalized genuine multiparty negativity (GMN)~\cite{Jungnitsch2011,Hofmann2014} to apply to fermionic systems. This measure of fermionic GMN (fGMN) retains analogous desirable qualities to the original renormalized GMN, such as being an entanglement monotone, reducing to a semidefinite programming (SDP) optimization problem, and being directly comparable to the minimal bipartite negativity in many situations. In Section~\ref{sec:preliminaries}, we discuss the previous entanglement measures that motivate the specific form of the fGMN. This form will be described in Section~\ref{section:fgmn_measure}, where we will also provide its various properties. In Section~\ref{sec:numerics}, we perform numerical tests of this new measure, showing the main observational differences between the fermionic GMN and its non-fermionic counterpart, as well as between multiparty and bipartite entanglement. 

\section{Preliminaries}~\label{sec:preliminaries}

We will start by reviewing entanglement. A quantum state $\rho$ on a bipartite system $A_1|A_2$ is entangled if it cannot be written in the form of a separable state
\begin{gather}
    \rho_{\text{sep}} = \sum_k p_k \rho^k_{A_1} \otimes \rho^k_{A_2}
\end{gather}
where $p_k$ are non-negative mixing coefficients that sum to $1$. A useful measure of bipartite entanglement is the entanglement negativity~\cite{Horodecki1996,Peres1996}
\begin{gather}
    \mathcal{N}(\rho) = \half\tr(|\rho^{T_1}|) - \half
\end{gather}
where $T_1$ is a partial transpose along subregion $A_1$ and $|\rho^{T_1}| = \sqrt{\rho^{T_1} (\rho^{T_1})^\dagger}$, such that $\tr(|\rho^{T_1}|)$ is the trace norm, or sum of singular values, of $\rho^{T_1}$. A typical state that is entangled between the subregions will acquire negative eigenvalues under the partial transpose, while retaining unit trace, leading to a sum of singular values greater than 1. 

The negativity is an entanglement monotone~\cite{Vidal2000}, which entails many useful properties. Firstly, while the negativity does not give a positive quantity for all entangled states~\cite{Horodecki1998}, it does evaluate to zero for all unentangled states. Secondly, it is monotonic under local operations and classical communications (LOCCs) - that is, no LOCCs applied to the state can increase the entanglement. Finally, it is upper bounded by $\half (d_{\text{min}}-1)$, where $d_{\text{min}}$ is the minimum dimension of either party. Aside from being a monotone, the negativity, in its logarithmic form $\log \tr(|\rho^{T_1}|)$, is additive and upper bounds the distillable entanglement in a bipartite state~\cite{Vidal2002}.

For the multipartite case, we should not only exclude unentangled states, but also states with incomplete entanglement, where a subregion might be entangled to some parties but not others. For example, we can consider product states of the form $\rho_m \otimes \rho_{\overline{m}}$, where $m|\overline{m}$ forms a partition of all the parties into two sets. Such a state might have entanglement between parties within one side of the partition, but not across the partition. The convex hull of all such states, over all partitions $m|\overline{m}$, forms the set of biseparable states,
\begin{gather}
    \rho_{\text{bisep}} = \sum_k \sum_m p^k_m \, \rho^k_m \otimes \rho^k_{\overline{m}}
\end{gather}
where $k$ is a mixing index. A state has genuine multiparty entanglement (GME) if it cannot be written in this biseparable form.

\subsection{Fermionic Entanglement}

Entanglement in fermionic systems differs from bosonic systems or systems of distinguishable particles, due to fermion parity superselection~\cite{Moriya2006,Banuls2007,Friis2013,Szalay2021,Vidal2021}. This restricts the LOCC class to operations which do not locally violate fermion parity (called fermionic LOCC, or fLOCC). This restricted class of allowed operations mean that some states which could be created from product states by LOCC cannot be created by fLOCC, and are therefore fermionically entangled, leading to more robust entanglement in general. 

Fermionic exchange statistics become important when trying to generalize the bipartite negativity to fermionic systems, as the partial transpose can potentially exchange fermion operators. This leads to the form of the fermionic partial transpose (FPT) introduced by Ref.~\onlinecite{Shapourian2017}, based on performing a partial time reversal on the target subregion. For a generic even fermionic operator $\mathcal{O}$ on a bipartite system $\{i_1 ... i_A\}|\{j_1 ...j_B\}$ expressed in terms of Majorana fermions $\gamma_{2i-1} = c_i + c_i^{\dagger}, \gamma_{2i} = i(c_i-c_i^{\dagger})$,
\begin{gather}
    \mathcal{O} = \sum_{\vec{a},\vec{b} \in \{0,1\}^{A,B}} \mathcal{O}_{\vec{a}\vec{b}} \; \gamma_{2i_1-1}^{a_1} \gamma_{2i_1}^{a_1'} ... \gamma_{2i_A-1}^{a_A}\gamma_{2i_A}^{a_A'} \gamma_{2j_1-1}^{b_1} ...\gamma_{2j_B}^{b_B'}
\end{gather}
the partial time reversal on subsystem $A$ takes the form
\begin{gather}
    \mathcal{O}^{R_A} = \sum_{\vec{a},\vec{b} \in \{0,1\}^{A,B}} \mathcal{O}_{\vec{a}\vec{b}} \; i^{|\vec{a}'|-|\vec{a}|}\gamma_{2i_1-1}^{a_1}... \gamma_{2i_A}^{a_A'} \gamma_{2j_1-1}^{b_1} ...\gamma_{2j_B}^{b_B'},
\end{gather}
that is, every odd Majorana fermion $\gamma_{2i_\ell-1}$ in $A$ acquires a phase of $-i$, while every even Majorana fermion $\gamma_{2i_\ell}$ in $A$ acquires a phase of $+i$. 
The fermionic negativity of a bipartite state $\rho$ is then the trace norm of $\rho$ under the fermionic partial transpose. This is, in general, the most appropriate way to convert the original partial transpose, as the resulting measure is an additive entanglement monotone~\cite{Shapourian2019}. 
Unlike the original partial transpose, the partial time reversal can potentially make the state non-Hermitian with complex eigenvalues~\cite{Matsuda2025,Xu2025}, and it is this complexity that provides the greater-than-unit trace norm which leads to finite entanglement. 

Operationally, fermionic entanglement can be seen as a quantum resource that is required to perform specific tasks under fLOCC. For example, we can consider a decoding or partial tomography problem where one has to identify the phase $s = \pm 1$ in the bipartite state
\begin{gather}
    \rho_{AB} = \frac{1}{4}\left(I + is \gamma_{A_1} \gamma_{A_2}\right)
\end{gather}
where $\gamma_{A_1}, \gamma_{A_2}$ are two Majorana fermions local to parties $A_1,A_2$ respectively. This phase cannot be identified by parties $A_1,A_2$ under fLOCC alone, without a fermionically entangled resource.

This problem can be extended to multiple parties. Starting with the three-party state
\begin{gather}
    \rho_{ABC} = \frac{1}{8}(I+s_1g_1)(I+s_2g_2)
\end{gather}
where $g_1, g_2$ are quartic Majorana operators
\begin{gather}
    g_1 = -\gamma_{A_1}^1\gamma_{A_1}^2\gamma_{A_2}^1\gamma_{A_3}^2\\
    g_2 = -\gamma_{A_1}^2\gamma_{A_2}^1\gamma_{A_2}^2\gamma_{A_3}^1\\
    g_1 g_2 = -\gamma_{A_1}^1\gamma_{A_2}^2\gamma_{A_3}^1\gamma_{A_3}^2
\end{gather}
and $\gamma_{A_k}^j, j\in\{1,2\}$ are Majorana fermions of flavor $j$ local to party $A_k$. Here, the phases $s_1, s_2 = \pm 1$ can only be identified by parties $A_1,A_2,A_3$ if they have access to a resource state with fermionic GME - no single fermionic biseparable resource can identify the phases with fLOCC. See \supp Appendix~\ref{app:decoding} for proofs of the decodability of these states.

\subsection{Genuine Multiparty Negativity}

The genuine multiparty negativity (GMN)~\cite{Jungnitsch2011} is an efficiently computable entanglement monotone that generalizes entanglement negativity to multiple parties and can detect a large set of entangled states. The  renormalized~\cite{Hofmann2014} definition of GMN is
\begin{gather}\label{eq:bosonic_gmn}
    \mathcal{N} = -\inf_{W \in \mathcal{W}} \tr (\rho W)
\end{gather}
where $\mathcal{W}$ is the set of all fully decomposable witnesses $W$: for all bipartitions $m|\overline{m}$ of the subsystem, there must exist matrices $P_m, Q_m$ such that
\begin{align}
    &W = P_m + Q_m^{T_m}\label{eq:sum_consistency_condition}\\
    &0 \preceq P_m\\
    &0 \preceq Q_m \preceq I
\end{align}
Here $A \preceq B$ indicates that $B-A$ is positive semidefinite. Heuristically, the $Q_m$ matrix forms the part of $W$ that detects the negativity along the bipartition $m|\overline{m}$ - in the bipartite case for example, $Q_m$ is optimal when it projects onto all negative eigenstates of $\rho^{T_m}$. The $P_m$ matrix is a positive ``buffer" which only increases $\tr(\rho W)$, but makes (\ref{eq:sum_consistency_condition}) consistent over each bipartition.
This measure is an entanglement monotone like the original bipartite negativity, and is in fact equal to the bipartite negativity when evaluated on a 2-party system. It is also more tractable than a typical multiparty entanglement measure, as the optimization over $\mathcal{W}$ takes the form of a semidefinite programming (SDP) problem, a class of optimization problems that are particularly efficient to solve. 

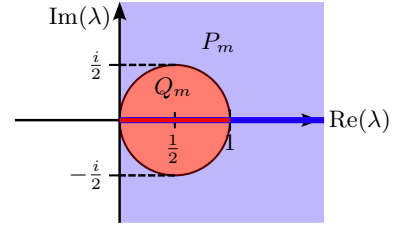
\begin{figure}[h]
    \centering
    \begin{tikzpicture}
        \begin{scope}
            \node[anchor=north west,inner sep=0] (image_a) at (0,0)
            {\includegraphics[width=0.22\textwidth]{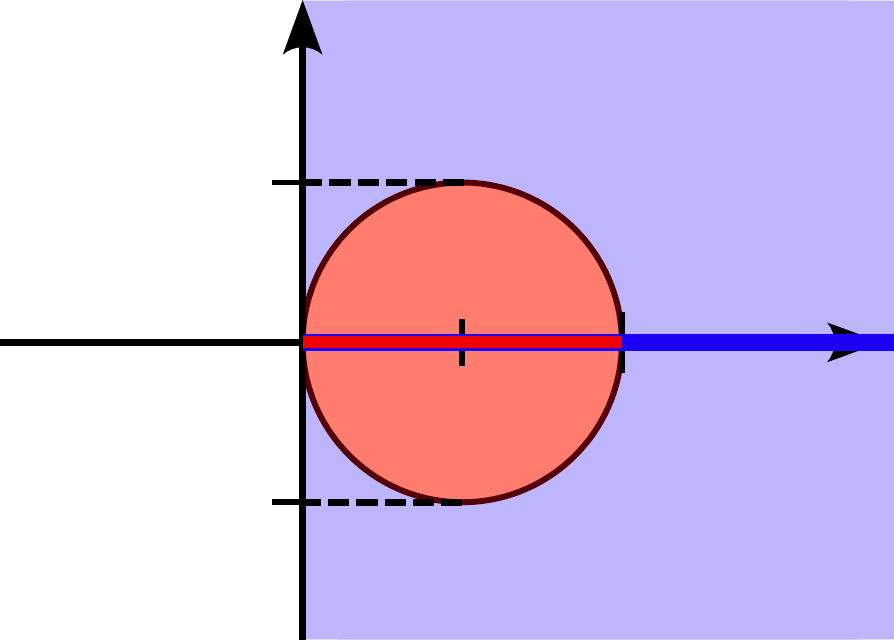}};
            \node [anchor=south east] (note) at (1.4,-0.5) {Im($\lambda$)};
            \node [anchor=south east] (note) at (5.1,-1.86) {Re($\lambda$)};
            \node [anchor=south east] (note) at (1.3,-1.15) {\small{$\frac{i}{2}$}};
            \node [anchor=south east] (note) at (1.3,-2.6) {\small{$-\frac{i}{2}$}};
            \node [anchor=south] (note) at (2.1,-1.4) {$Q_m$};
            \node [anchor=north] (note) at (2.1,-1.65) {\small{$\frac{1}{2}$}};
            \node [anchor=north] (note) at (2.85, -1.65) {$1$};
            \node [anchor=north] (note) at (2.7, -0.3) {$P_m$};
        \end{scope}
    \end{tikzpicture}
    \vspace*{-0.3cm}
    \caption{Allowed eigenvalues $\lambda$ for the $Q_m$ and $P_m$ matrices on the complex plane. The solid red and blue lines are the allowed eigenvalues for $Q_m$ and $P_m$, respectively, in the non-fermionic case, which must be real and positive. The red and blue shaded areas are the allowed eigenvalues for $Q_m$ and $P_m$ in the fermionic case, respectively, which are now allowed to have imaginary components.}
    
    \label{fig:fgmn_allowed_eigenvalues}
\end{figure}

We will now use the non-fermionic GMN (\ref{eq:bosonic_gmn}) as inspiration to build a properly fermionic alternative. The first step will be to replace the partial transpose $Q_m^{T_m}$ with the fermionic partial transpose $Q_m^{R_m}$ of Ref.~\onlinecite{Shapourian2017}. 
This by itself will not recover the bipartite fermionic negativity in the 2-party case. In this case, there is only one partition $m|\overline{m} = 1|2$; therefore, the positive buffer $P_1$ is zero at the infimum, giving
\begin{align*}
    \mathcal{N}^F = -\inf_{Q_1} \tr(\rho Q_1^{R_1})= -\inf_{Q_1} \tr(\rho^{R_1} Q_1)
\end{align*}
where we used the self-adjoint nature of the fermionic partial transpose (\supp Appendix~\ref{app:fpt_identities}).
Because $\rho^{R_1}$ is still normal, it admits an eigenvalue decomposition with complex eigenvalues $\lambda_i$, so the negativity can be written as $\half(\sum_i |\lambda_i|-1)$. We recover this sum if we require $Q_1$ to have the same eigenstates as $\rho^{R_1}$, with eigenvalues 
\begin{align}
    \mu_i = \frac{1}{2}\big(1-\frac{\lambda_i^*}{|\lambda_i|}\big)
\end{align}
(We set $\mu_i = 0$ whenever $\lambda_i = 0$). This suggests that the eigenvalues of $Q_m$ should lie in the complex disk \{$\frac{1}{2}+z, |z| \leq \frac{1}{2}$\} (Fig.~\ref{fig:fgmn_allowed_eigenvalues}), which relaxes the SDP restrictions $0 \preceq Q_m \preceq I$ to an operator norm restriction $||2Q_m-I|| \leq 1$. Formally this is a second-order cone condition, equivalent to requiring that the matrix
\begin{align}
    \begin{pmatrix}I & (2Q_m-I) \\ (2Q_m-I)^\dagger & I\end{pmatrix}
\end{align}
is positive semidefinite - therefore, the condition still keeps the problem in the SDP class.

Because we have relaxed the Hermitian condition on $Q_m$, we also allow $P_m$ to contain an arbitrary anti-Hermitian component, while still requiring that its Hermitian part is positive semidefinite ($P_m + P_m^\dagger \succeq 0$). This allowance will be important for proving analogous results to the renormalized GMN from Ref.~\onlinecite{Hofmann2014}. As a consequence, our negativity will only optimize the real part of $\tr(\rho W)$. For $P_m + P_m^\dagger \succeq 0$, we have $\re \tr(\rho P_m) \geq 0$ for all states $\rho$, so the role of $P_m$ remains as a positive buffer to reduce negativity.

\section{The Fermionic GMN}~\label{section:fgmn_measure}

We propose the following definition of fermionic genuine multiparty negativity:
\begin{gather}
    \mathcal{N}^F = -\inf_{W \in \mathcal{W}^F} \re \tr (\rho W)
\end{gather}
where $\mathcal{W}^F$ is the set of all witnesses $W$ fully fermion-decomposable among all partitions: for all bipartitions $m|\overline{m}$ of the subsystem, there must exist $P_m, Q_m$ such that
\begin{align}
    &W = P_m + Q_m^{R_m}\label{eq:fermion_decomposable_condition_1}\\
    &P_m + P_m^\dagger \succeq 0 \label{eq:fermion_decomposable_condition_2}\\
    &||2Q_m-I|| \leq 1 \label{eq:fermion_decomposable_condition_3}
\end{align}
We will now prove that this negativity has the following properties:
\begin{description}
    \item[\textit{(i)}] $\mathcal{N}^F(\rho) = 0$ for all fermion-biseparable states (i.e. convex combinations of $\rho_m \otimes \rho_{\overline{m}}$, where $\rho_m$ and $\rho_{\overline{m}}$ are both fermion parity preserving).
    \item[\textit{(ii)}] $\mathcal{N}^F(\rho)$ is monotonically decreasing under fLOCC, and invariant under local basis change.
    \item[\textit{(iii)}] $\mathcal{N}^F(\rho)$ is convex: $\mathcal{N}^F(\sum_k p_k\rho_k) \leq \sum_k p_k \mathcal{N}^F(\rho_k)$ for all decompositions of $\rho$ into states $\rho_k$ with positive coefficients $p_k$.
    \item[\textit{(iv)}] $\mathcal{N}^F(\rho)$ reduces to the bipartite fermion negativity in the 2-party case.
    \item[\textit{(v)}] $\mathcal{N}^F(\rho) \leq \min_m \mathcal{N}^F_m(\rho)$, where $m$ indexes all possible nontrivial bipartitions of the parties and $\mathcal{N}^F_m$ is the fermionic bipartite negativity across the partition $m|\overline{m}$, with equality if $\rho$ is pure.
    \item[\textit{(vi)}] $\mathcal{N}^F(\rho) \leq \half (d_{\text{min}}-1)$, where $d_{\text{min}}$ is the minimum dimension of any party in the system.
\end{description}
\noindent\textit{Proof of (i)}: For a fermionic biseparable state,
\begin{gather}
    \rho = \sum_k \sum_m p_m^k \, \rho^k_m\otimes \rho^k_{\overline{m}}
\end{gather}
where $m$ is the index of partitions over the system, $k$ is a mixing index, $p_m^k$ are non-negative coefficients, and $\rho^k_m \otimes \rho^k_{\overline{m}}$ is a fermionic product state with $\rho^k_m, \rho^k_{\overline{m}}$ preserving fermion parity. So for any fully fermion-decomposable witness $W$,
\begin{align*}
    \re\tr(\rho W) &= \re \sum_{k,m} p_m^k \tr\big(\rho^k_m \otimes \rho^k_{\overline{m}} ( P_m + Q_m^{R_m})\big)\\
    &\geq\re \sum_{k,m} p_m^k \tr\big((\rho^k_m \otimes \rho^k_{\overline{m}})  Q_m^{R_m}\big)\\
    &=\re \sum_{k,m} p_m^k \tr\big[\big((\rho^k_m)^{R} \otimes \rho^k_{\overline{m}}\big)  Q_m\big]\\
    &=\half + \half \re \sum_{k,m} p_m^k \tr\big[\big((\rho^k_m)^{R} \otimes \rho^k_{\overline{m}}\big)  (2Q_m-I)\big]
\end{align*}
Because $\rho^k_m$ respects overall fermion parity, it remains a Hermitian, positive semidefinite operator under the fermionic transpose (\supp Appendix~\ref{app:fpt_identities}). Therefore we can decompose $\rho^k_m \otimes \rho^k_{\overline{m}}$ into an eigenbasis $\sum_j r^k_{mj} \ket{\phi^k_{mj}}\bra{\phi^k_{mj}}$, with $r^k_{mj} \geq 0, \sum_j r^k_{mj} = 1$ (here $\ket{\phi^k_{mj}}$ is a member of the whole Hilbert space over $m|\overline{m}$), giving
\begin{align}
    \re\tr(\rho W) &\geq \half \! + \! \half \sum_{k,m} p^k_m \sum_j r^k_{mj} \re\! [\bra{\phi^k_{mj}} (2Q_m\!-\!I) \ket{\phi^k_{mj}}] \n
    &\geq \half - \half \sum_{k,m} p^k_m \sum_j r^k_{mj} \left|\bra{\phi^k_{mj}} (2Q_m-I) \ket{\phi^k_{mj}}\right|\n
    &\geq \half - \half \sum_{k,m} p^k_m \sum_j r^k_{mj} = 0,
\end{align}
using the non-negativity of all $p^k_m, r^k_{mj}$ coefficients in the second line and the operator norm bound $||2Q_m - I|| \leq 1$ in the third. Hence, any valid witness $W$ must have $\re \tr(\rho W) \geq 0$ for fermionic biseparable $\rho$, with equality at least at $W=0$, giving $\mathcal{N}^F = -\inf_W \re \tr(\rho W) = 0$. \ep

\noindent\textit{Proof of (ii)}: A generic fLOCC channel $\Lambda(\rho) = \sum_i K_i \rho K_i^\dagger$ has Kraus components $K_i = K^1_i \otimes K^2_i \otimes ... \otimes K^P_i$ which are separable into local, parity-conserving terms. 
Given this, our inequality is proven as long as $\Lambda^\dagger(W)$ remains in $\mathcal{W}^F$, where $\Lambda^\dagger = \sum_i K_i^\dagger (.) K_i$ is the adjoint channel. We remain in $\mathcal{W}^F$ provided 
\begin{gather}
    \Lambda^\dagger(P_m) + [\Lambda^\dagger(P_m)]^\dagger = \Lambda^\dagger(P_m+P_m^\dagger)
\end{gather}
is still positive semidefinite, and that $\widetilde{Q}_m = [\Lambda^\dagger(Q_m^{R_m})]^{R_m^*}$ still obeys $||2\widetilde{Q}_m -I|| \leq 1$, where $R_m^*$ is the inverse, or complex conjugate, fermionic partial transpose (see \supp Appendix~\ref{app:fpt_identities}). The former is immediately given from $\Lambda$ being a completely positive map. For the latter, we have
\begin{align*}
    \widetilde{Q}_m 
    &= \bigg[\sum_i (L_i^\dagger \otimes M_i^\dagger) Q_m^{R_m} (L_i \otimes M_i) \bigg]^{R_m^*}
\end{align*}
where $M_i$ acts solely on the subregions of $m$ and $L_i$ acts solely on the complement. Proposition 2 of Ref.~\onlinecite{Shapourian2019} (restated in \supp Appendix~\ref{app:fpt_identities}) allows us to distribute the outer fermionic partial transpose among the individual factors: 
\begin{align*}
    \widetilde{Q}_m 
    &= \sum_i (L_i^\dagger \otimes M_i^{R^*}) Q_m (L_i \otimes (M_i^{R})^\dagger)
\end{align*}
Now, $M_i$ preserves fermion parity on $m$, so $M_i^{R} = M_i^{R^*}$ is some unitary matrix $N_i^\dagger$, with $(M_i^{R})^\dagger = N_i$. Therefore, $\widetilde{Q}_m$ is just $Q_m$ under a completely positive unital channel, and the operator norm bound $||2Q_m -I|| \leq 1$ still holds for $\widetilde{Q}_m$. \ep

\noindent\textit{Proof of (iii)}: The proof is identical to the non-fermionic case~\cite{Jungnitsch2011,Hofmann2014}. For a convex mixture of states $\rho = \sum_k p_k \rho_k$,
\begin{align*}
    \mathcal{N}^F(\rho) 
    &= -\inf_{W \in \mathcal{W}^F} \sum_k p_k \re \tr(\rho_k W)\\
    &\leq - \sum_k p_k \inf_{W \in \mathcal{W}^F} \re \tr(\rho_k W) =  - \sum_k p_k \mathcal{N}^F(\rho_k),
\end{align*}
with equality if some $W$ is the optimal witness for both $\rho$ and each individual $\rho_k$. \ep 

\noindent\textit{Proof of (iv)}: For the 2-party case, there is only one partition, with $W = P_1 + Q_1^{R_1}$. Because $\re \tr(\rho P_1) \geq 0$, it remains optimal to set that component to zero, and just optimize over $Q_1$. Taking the singular value decomposition $\rho^{R_1} = \sum_i s_i \ket{u_i}\bra{v_i}$, this gives
\begin{align}
    \mathcal{N}^F 
    &=-\half - \half\inf_{Q_1} \sum_i s_i \re \bra{v_i} (2Q_1-I) \ket{u_i}\n
    &\leq-\half + \half\sup_{Q_1}\sum_i s_i ||(2Q_1-I) \ket{u_i}||\n
    &\leq-\half + \half\sum_i s_i 
    =\half \tr(|\rho^{R_1}|) - \half,
\end{align}
which is the original definition of bipartite fermionic negativity, thus $\mathcal{N}^F$ is upper bounded by it. Our bound reaches equality at $2Q_1 - I = -\sum_i \ket{v_i} \bra{u_i}$, which still respects the operator norm restriction on $2Q_1 - I$, so the two definitions of bipartite fermionic negativity must in fact be equal. \ep

\noindent\textit{Proof of (v),(vi)}: This proof depends on the dual of the SDP problem. The dual is derived in \supp Appendix~\ref{app:dual_sdp}, and takes the form
\begin{gather}
    \mathcal{N}^F(\rho) = \inf\bigg\{ \sum_m p_m \mathcal{N}^F_m(\rho_m)\,\bigg|\, \rho_m \succeq 0 \,\, \forall \, m, \n \qquad\qquad\qquad\qquad\text{ and } \rho = \sum_m p_m \rho_m\bigg\}\label{eq:dual_sdp_1}\\
    \mathcal{N}^F_m(\rho_m) = \half \tr(|\rho_m^{R_m}|) - \half \label{eq:dual_sdp_2}
\end{gather}
where $m$ indexes valid partitions of the parties. We will use $\inf_{\rho = \sum_m p_m \rho_m}$ as a shorthand for the infimum over all set conditions of (\ref{eq:dual_sdp_1}).
The proof of \textit{(v)} follows as $\mathcal{N}^F_m$, from (\ref{eq:dual_sdp_2}), is simply the fermion bipartite negativity across the partition $m|\overline{m}$, and
\begin{align*}
    \mathcal{N}^F(\rho) &= \inf_{\rho=\sum_m p_m \rho_m} \sum_m p_m \mathcal{N}^F_{m}(\rho_m) \\
    &\leq \mathcal{N}^F_m(\rho) \qquad \forall m
\end{align*} using the trivial decomposition $\rho = \rho_m$ for some $m$, which is the only possible decomposition for a pure state.

As a corollary, property \textit{(vi)} follows from property \textit{(v)} and the previously established bound $\mathcal{N}^F_m \leq \half(\min\{d_m, d_{\overline{m}}\}-1)$ for the bipartite negativity~\cite{Shapourian2019}. \ep

\subsection{Minimum bipartite negativity}
Let us review Property \textit{(v)} of the fermionic GMN - that, for pure states, it is equal to the minimum fermionic bipartite negativity across any nontrivial partition of the parties:
\begin{gather}
    \mathcal{N}^F(\ket{\psi}\bra{\psi}) = \min_m \mathcal{N}^F_m(\ket{\psi}\bra{\psi})
\end{gather}
It turns out that pure states are not the only class of states with this property. We can make the following addition:
\begin{description}
    \item[\textit{(v.b)}] For states of the form
    \begin{gather}\label{eq:mixed_stabilizer_state}
        \rho = 2^{-(N-S)} \prod_{s=1}^S \half\big(I+g_s\big),
    \end{gather}
    where $g_s$ are generators of a fermionic stabilizer group, $\mathcal{N}^F(\rho) = \min_m \mathcal{N}^F_m(\rho)$.
\end{description}
This form describes mixed stabilizer states - the maximal mixture over all simultaneous $+1$ eigenstates of each stabilizer generator.
The proof of Property \textit{(v.b)} is in \supp Appendix~\ref{app:stabilizer_fgmn} - instead of converting to the SDP dual, we use the stabilizer generators to define channels which simplify the problem without changing the optimum.

Some examples of these states include ground states of frustration-free fermion Hamiltonians, such as Majorana dimer states, graph states of even degree, and code states in Majorana fermion codes~\cite{Bravyi2010}. 
For a specific 3-party example, we can consider the state
\begin{gather}
    \rho = 2^{-3}(I-\gamma_1 \gamma_2 \gamma_3 \gamma_6)(I-\gamma_2 \gamma_3 \gamma_4 \gamma_5)
\end{gather}
where the fermionic modes corresponding to each party are $c_j = \half(\gamma_{2j-1} - i\gamma_{2j})$. For any partition of the three sites, there is always at least one stabilizer in the group with an odd number of fermions on one side of the partition - therefore, $\min_m \mathcal{N}^F_m$ and hence $\mathcal{N}^F$ are nonzero. However, there is no non-fermionic GME, as our state only has two stabilizers. This example highlights the significantly relaxed conditions that fermionic GME requires, compared to its non-fermionic counterpart.

\subsection{Lower bounds on fermionic GMN}
As described in Ref.~\onlinecite{Jungnitsch2011}, the nature of the GMN as an optimization problem over a large set of witnesses means that optimizing over any subset of the witnesses gives a lower bound on the GMN. This property holds for the fermionic measure as well. In particular, if the density matrix is not known, and we instead have observables $O_i$ with measured expectation values $q_i$ over the state, we can find a lower bound without any further measurements by optimizing over all fully fermion-decomposable linear combinations of observables, 
\begin{gather}
    W  = \sum_i \lambda_i O_i \in \mathcal{W}\\
    \tr(\rho W) = \sum_i \lambda_i q_i.
\end{gather}
As more observables are measured, the set of such witnesses expands, and the lower bound will become closer to the true optimum.

\section{Evaluating fermionic GMN}\label{sec:numerics}

\subsection{White noise mixed states}

\begin{figure}[h]
    \centering
    \begin{tikzpicture}
        \begin{scope}
            \node[anchor=north west,inner sep=0] (image_a) at (0,0)
            {\includegraphics[width=0.53\textwidth]{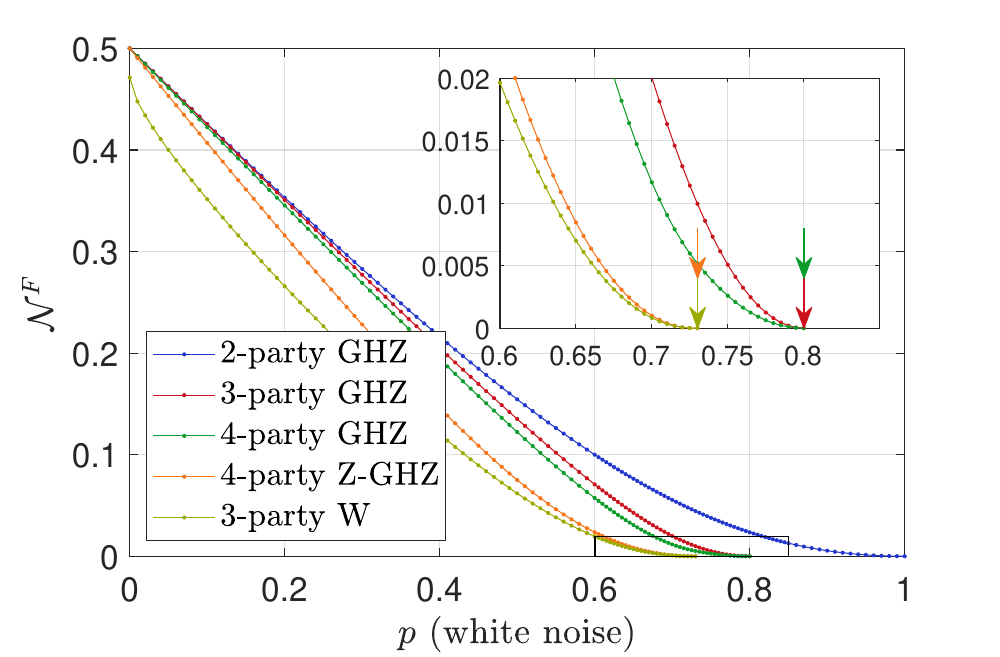}};
        \end{scope}
    \end{tikzpicture}
    \vspace*{-0.3cm}
    \caption{Fermionic GMN for GHZ and W states under white noise mixing. The party of entanglement measured is always the number of qubits in the state. 
    Sudden death points of the multiparty mixed states, where the GMN drops to zero, are indicated by arrows in the inset.}
    
    \label{fig:white_noise_mixed_states}
\end{figure}

\begin{table}[]
    \centering
    \begin{tabular}{|c| c|c |} \hline
        State & \begin{tabular}{@{}c@{}}Sudden death\\mixing\end{tabular} & \begin{tabular}{@{}c@{}}Lower bound\\sudden death\end{tabular} \\ \hline 
        GHZ$_2$,X & 1 & 1\\ \hline 
        GHZ$_3$,X & 0.799962 & 0.8\\ \hline 
        GHZ$_4$,X & 0.799949 & 0.8\\ \hline
        GHZ$_4$,Z & 0.727227 & 8/11\\ \hline
        W$_3$ & 0.727224 & 8/11\\ \hline
    \end{tabular}
    \caption{Sudden death mixing probabilities for the 2-4 qubit X-basis GHZ, 4-qubit Z-basis GHZ and 3-qubit W state, according to the fermionic GMN. Lower bound sudden death points indicate white noise mixings where a fermionic biseparable mixture can be explicitly produced (\supp Appendix~\ref{app:bisep_forms}).}
    \label{tab:white_noise_sudden_deaths}
\end{table}

We first test our fermionic GMN on simple pure states $\ket{\Psi}$ under white noise mixing of strength $p$,
\begin{gather}
    \rho = (1-p)\ket{\Psi}\bra{\Psi} +  p\frac{I}{d}\,.
    \end{gather}
In particular, we will look at 2-4 qubit GHZ states in the Pauli X-basis, the 4 qubit GHZ in the Pauli Z-basis, and 3 qubit W state under white noise. 

The behavior of the fermionic GMN under this mixing is recorded in Figure~\ref{fig:white_noise_mixed_states}. For two parties, it is possible to have finite fermionic entanglement at any white noise level $p < 1$, which is recovered for our more generic measure. However, higher party entanglement can experience a sudden death mixing $p^*$, beyond which no entanglement is measured at all. This is a natural consequence of fermionic biseparable states forming a finite-measure subset of all fermionic states~\cite{Parez2026}, as opposed to fermionic separable states, which form a measure zero subset. In fact, for most of these states, it is possible to find a valid fermionic biseparable decomposition at a level of mixing almost at the sudden death point reported by the GMN (Table~\ref{tab:white_noise_sudden_deaths}).

\subsection{Thermal states}

We can also test our fermionic GMN on more physically motivated mixed states, such as thermal states of free fermion models like the one-dimensional Kitaev chain
\begin{gather}
    H = -\sum_i t f^{\dagger}_{i} f_{i+1} + \Delta f_i^{\dagger} f_{i+1}^\dagger + \text{h.c.} - \sum_i \mu n_i
\end{gather}
For $t=\Delta$, there exists a phase transition at $\mu/t = 2$ from a topologically nontrivial phase containing Majorana edge modes with an exponentially suppressed energy gap~\cite{Kitaev2001,Chiu2016}, to a trivial phase. Because the system is non-interacting, the states are Gaussian and described entirely by their 2-body correlation functions, allowing us to simulate large systems efficiently. Unlike the bipartite case, however, the sizes of the reduced density matrices are still constrained by the entanglement measure, whose SDP optimization algorithm can only be evaluated using RDMs in an occupation number basis. Therefore, we will measure entanglement in this system by first producing the covariance matrices of ground or thermal states, then, after restricting the matrices to the subregions of interest, convert to the occupation number basis using a Jordan-Wigner transform.

\begin{figure}[h]
    \centering
    \begin{tikzpicture}
        \begin{scope}
            \node[anchor=north west,inner sep=0] (image_a) at (0,0)
            {\includegraphics[width=0.47\textwidth]{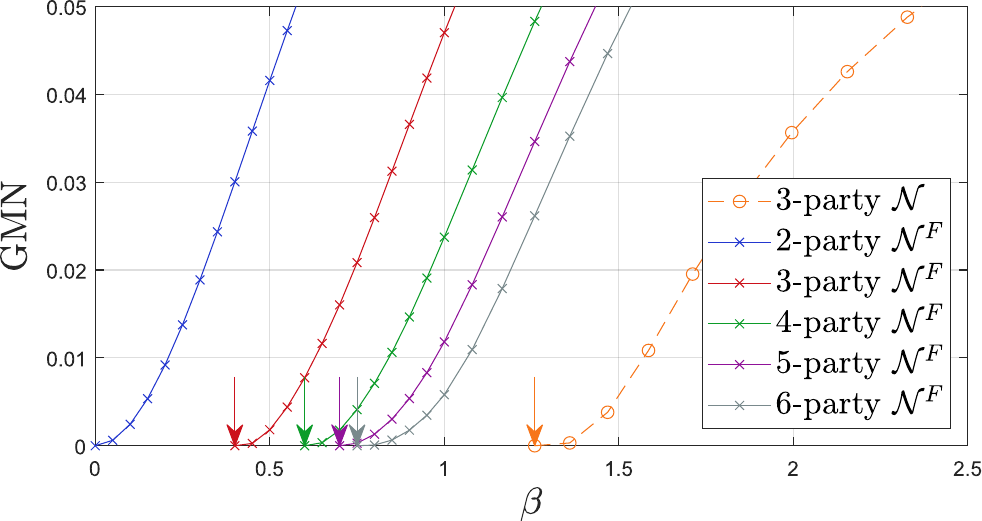}};
        \end{scope}
    \end{tikzpicture}
    \vspace*{-0.4cm}
    \caption{Fermionic and non-fermionic GMN for different configurations in a $N=1000$ site Kitaev chain with $t=\Delta=1, \mu=2$. 
    Configurations are 2-6 sites in the center of the chain, with each subregion being one qubit. Arrows indicate sudden death points.}
    \label{fig:thermal_kitaev}
\end{figure}

\begin{table}[]
    \centering
    \begin{tabular}{|c||c|c|} \hline
        Parties & \begin{tabular}{@{}c@{}}Sudden death $\beta$\\(non-fermionic)\end{tabular}& \begin{tabular}{@{}c@{}}Sudden death $\beta$\\(fermionic)\end{tabular}\\ \hline 
        2 & 0.9661 & 0 \\ \hline
        3 & 1.3425 & 0.4204 \\ \hline
        4 & 1.5698 & 0.6087 \\ \hline
        5 & 1.7424 & 0.7041\\ \hline
        6 & 1.8683 & 0.7701 \\ \hline
    \end{tabular}
    \caption{Sudden death temperatures for 2-6 qubit non-fermionic and fermionic negativities of the system in Fig.~\ref{fig:thermal_kitaev}.}
    \label{tab:thermal_sudden_deaths}
\end{table}

In Figure~\ref{fig:thermal_kitaev}, we compare the fermionic GMN for a collection of 2-6 sites in the center of a 1000-site chain, at the critical point of the Kitaev model. The bipartite fermionic GMN is the only entanglement measure without a sudden death temperature - similar to the white-noise mixing, fermionic GMN at three or more parties experience a temperature beyond which no entanglement is detected (Table~\ref{tab:thermal_sudden_deaths}).

\subsection{Pure states} \label{sec:kitaev_ground_state}

\begin{figure}[h]
    \centering
    \begin{tikzpicture}
        \begin{scope}
            \node[anchor=north west,inner sep=0] (image_a) at (-0.03,0)
            {\includegraphics[width=0.465\textwidth]{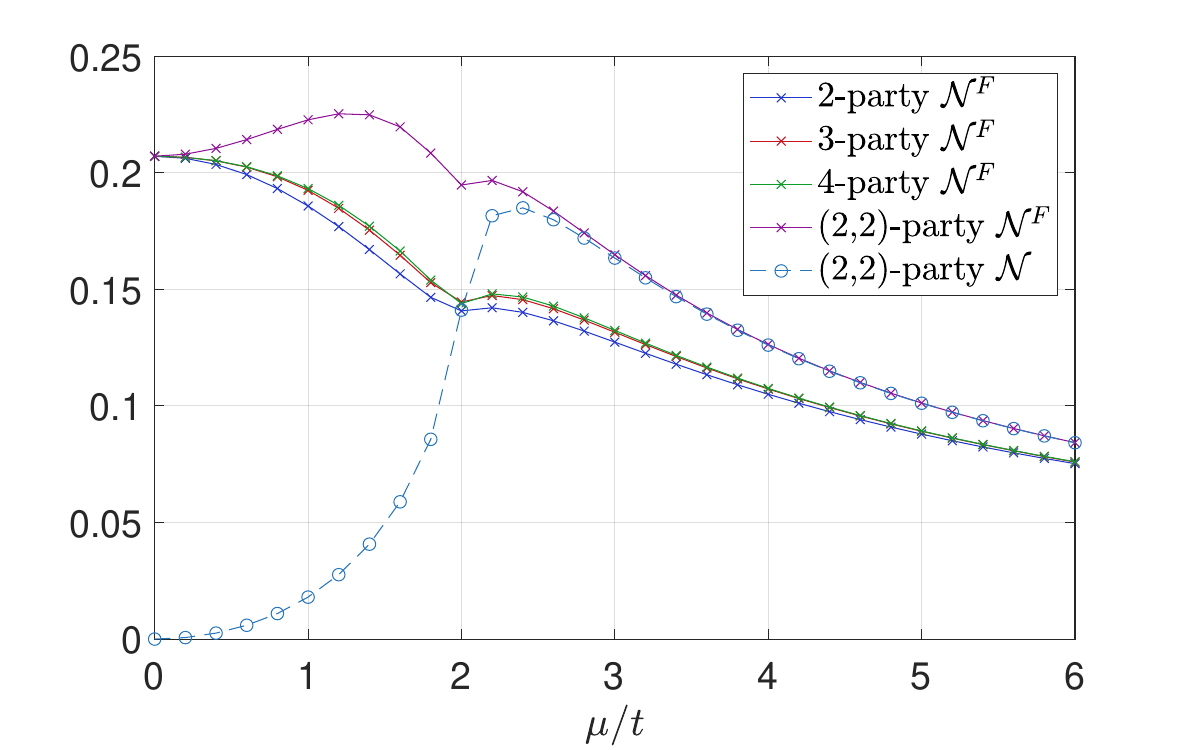}};
            \node [anchor=north west] (note) at (0.05,-0.2) {\textbf{a)}};
        \end{scope}
        \begin{scope}[yshift=-0.26\textwidth]
            \node[anchor=north west,inner sep=0] (image_a) at (0,0)
            {\includegraphics[width=0.47\textwidth]{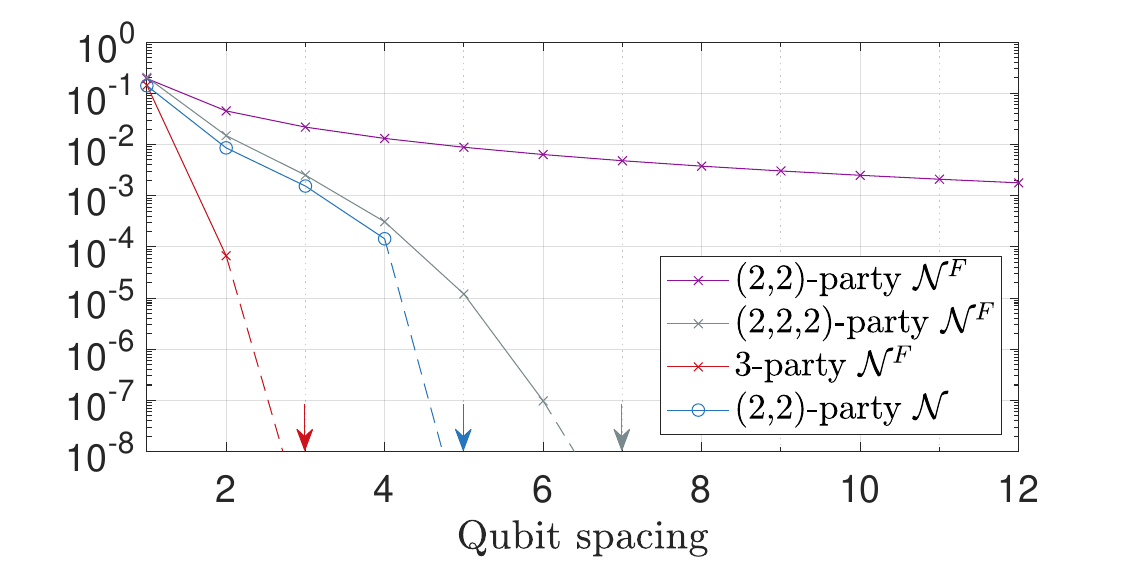}};
            \node [anchor=north west] (note) at (0.05,-0.15) {\textbf{b)}};
        \end{scope}
    \end{tikzpicture}
    \vspace*{-0.3cm}
    \caption{\textit{(a)}: Fermionic and non-fermionic GMN in the $N=1000$ Kiteav model ground state over $\mu/t$. Configurations are 2-4 qubits in the center of the chain, with each subregion being one qubit, except for the 22-party GMNs which measure over two subregions of two qubits each. \textit{(b)}: Fermionic and bosonic GMN over spacing between subregions at the critical point $\mu=2$, with a spacing of 1 site denoting nearest-neighbor subregions, etc. 222-party GMN measures tripartite entanglement over three subregions of two qubits each. Arrows indicate sudden death points.}
    \label{fig:kitaev_gs_over_mu}
\end{figure}

Figure~\ref{fig:kitaev_gs_over_mu}a shows the fermionic multipartite entanglement for the ground state of the Kitaev chain over all $\mu$, on the 1000-site chain.  

Near the $\mu=0$ point, all fermionic negativities converge to $\half(\sqrt{2}-1)$, while all non-fermionic entanglement drops to zero. This is a consequence of Property \textit{(v.b)} of the fermionic GMN - in the limit of $\mu=0$, the ground state of the Kitaev chain is the Majorana dimer state, and its RDM over $k$ contiguous sites has the form
\begin{gather}
    \rho = 2^{-k} \prod_{j=1}^{k-1} (I-i\gamma_{2j}\gamma_{2j+1})
\end{gather} 
Because this is a mixed stabilizer state, Property \textit{(v.b)} holds and the fermionic GMN is simply the minimum bipartite negativity across all nontrivial partitions. 
For this state, the bipartite negativity across a partition depends on the minimum number of stabilizers ``cut" by that partition - those that have an odd number of Majorana fermions on either side of the partition. Since the minimum number of cuts over a linearly continguous subregion is one, the GMN is the same as the negativity of a single Majorana bond. 

At finite $\mu$, the 2-4 party fermionic entanglement values differ. These differences then describe non-dimer components of the entanglement, that is, the longer-ranged influences on entanglement that occur in the bulk of the system.

Besides a sudden death in the GMN over temperature, we can also investigate the sudden death over distance between parties on the pure ground state. These types of sudden death have been observed for many unitary critical systems using non-fermionic entanglement measures~\cite{Osterloh2002,Javanmard2018}, while for bipartite fermions tends to be long ranged with power law decay~\cite{Parez2024b,Blanchet2024}. Figure~\ref{fig:kitaev_gs_over_mu}b measures different types of entanglement over the number of sites that separate each subregions. Like with the thermal case, the 2-party fermionic entanglement is the only type of entanglement that is long ranged. For three parties, we get a sudden death in entanglement for single-qubit subregions once the qubits have gone beyond next-nearest neighbor, though two-qubit subregions are allowed to have finite entanglement for separations of up to 6 sites.

\section{Outlook}

We have devised a measure of genuine multiparty entanglement that brings the efficiency and comprehensiveness of GMN to fermionic systems. With this, we will be able to compare the fermionic and non-fermionic entanglement of many more states in the future. For most of the systems we have already investigated, the fermionic multipartite entanglement tended to be more abundant than non-fermionic entanglement, and was more robust to noise. However, some broad properties, such as the presence of sudden death points and long-ranged scaling in critical systems, remain the same. There were also exceptions where multiparty fermionic entanglement behaved significantly differently from non-fermionic counterparts, such as the ground state of the Kitaev chain in the topological phase.

Are there any other systems where this difference is apparent? For one possible situation, the ground and Gibbs states of most fermionic Hamiltonians have area-law or log-law bipartite entanglement~\cite{Eisert2010} - that is, the leading term in the entanglement between two adjacent subregions is proportional to the boundary between the subregions in gapped systems, or logarithmic in their size for critical systems. By Property \textit{(v)}, this must be an upper bound for the fermionic GMN between multiple parties, as well. 
The argument of Ref.~\onlinecite{Lyu2025b} implies that this bound should be generically saturated in the limit of large party size. It would be interesting to test this saturation, especially in a system where the fermionic bipartite entanglement has log-law scaling.

On the technical side, one question of interest is whether the fermionic GMN can be adapted to Gaussian states in a way that only requires operating on the correlation matrix of the state. Such an adaptation is feasible for the fermionic bipartite negativity~\cite{Shapourian2017}, and in fact was part of the justification for the form of the fermionic partial transpose, but encounters problems in the multiparty case due to the composite nature of the witness. As the sum of two Gaussian states is not necessarily Gaussian, it does not make sense to demand the witness $W$ and all its possible components $P_m, Q_m$ to be Gaussian at the same time. It should be noted that property \textit{(v)} of the fermionic GMN automatically sets an upper bound based on the fermionic bipartite negativity, however it would not be able to detect some important properties of the fermionic GMN, such as sudden death, in general cases.
Another question is whether there is any clear interpretation of the fermionic GMN, or other measure of fermionic GME, from a field theory perspective, like there is with bipartite negativity~\cite{Calabrese2012}. That is, we could look for an expression of the fermionic GMN using the moments of the reduced density matrix in a replica model. However, this question does not have a clear answer for any measure of GME, fermionic or not.

\section{Acknowledgements}
We would like to thank Gilles Parez and Shinsei Ryu for their valuable feedback on the paper. We acknowledge MOSEK~\cite{mosek} and YALMIP~\cite{yalmip} for the SDP optimization software used to calculate the GMN. The custom scripts we used to generate numerical simulations, not including MOSEK and YALMIP, are openly available \cite{data}. W.W.-K. is supported by a grant from the Foundation Courtois, a Chair of the Institut Courtois, a Discovery Grant from NSERC, and a Canada Research Chair.

\clearpage
\onecolumngrid

\begin{appendices}

\renewcommand{\thesubsection}{\arabic{subsection}}
\renewcommand{\theequation}{S\arabic{equation}}

\section{Decoding states using fermionic GME resources}\label{app:decoding}

We will consider the following class of decoding problem: a multipartite fermionic stabilizer state
\begin{gather}
    \rho_{A_1 ... A_P}(s_1,...,s_S) = \frac{1}{d}\prod_{j=1}^S(I+s_jg_j)
\end{gather}
is sent to parties $A_1 ... A_P$, where $g_j$ are fermionic stabilizers and $s_j=\pm 1$. The parties have access to a resource state $\sigma$ and can perform fLOCC actions, and must identify all phases $s_j$. 

We will start with a bipartite state $\rho_{A_1 A_2}$ with one stabilizer over the Majorana fermions $\gamma_{A_1}, \gamma_{A_2}$ in parties $A_1, A_2$ respectively:
\begin{gather}
    \rho_{A_1 A_2}(s) = \frac{1}{4}(I + si\gamma_{A_1}\gamma_{A_2}).
\end{gather}
If parties $A_1, A_2$ are able to decode the phase $s$ through a resource state $\sigma$ and fLOCC actions, then there exists some fermionic POVM $\{M_t\}$ such that
\begin{gather*}
    \tr\left(M_t (\rho(s) \otimes \sigma)\right) = \delta_{st}\\
    M_+ + M_- = I, \qquad M_{\pm} \succeq 0\\
    M_t = \sum_{i} M_{t,A_1}^{(i)} \otimes M_{t,A_2}^{(i)}
\end{gather*}
where the components $M_{t,A_j}^{(i)}$ are local to party $A_j$ and respect fermion parity there. Expanding $\rho(s)$, we require
\begin{gather}
    \sum_i \tr\left(\big(M_{t,A_1}^{(i)} \otimes M_{t,A_2}^{(i)}\big) \big(\frac{1}{4}(I + si\gamma_{A_1}\gamma_{A_2}) \otimes \sigma\big)\right) = \delta_{st}.
\end{gather}
In order for any term to have a nonzero trace, fermion parity needs to be preserved on both parties. Since $\rho(s)$ violates local fermion parity, $\sigma$ must violate fermion parity in the same way to preserve the overall product, which can only be satisfied if $\sigma$ also has fermionic entanglement between the parties. Such an entangled resource state could be a copy of $\rho(1)$ itself: using $\sigma = \frac{1}{4}(I+i\gamma_{A_1}' \gamma_{A_2}')$ and POVM $M_t = \frac{1}{2}(I+t\gamma_{A_1} \gamma_{A_1}'\gamma_{A_2}\gamma_{A_2}')$, we have $\tr\big(M_t (\rho(s) \otimes\sigma)\big) = \delta_{st}$.

This separation between entangled and unentangled resources can be extended to the multiparty case. We start with the multiparty state 
\begin{gather}
    \rho_{A_1 A_2 A_3}(s_1,s_2) = \frac{1}{8}\left(I+s_1g_1\right)\left(I + s_2 g_2\right) = \frac{1}{8}\left(I+s_1g_1 + s_2g_2 + s_1 s_2 g_1 g_2\right) \label{eq:form_of_tripartite_stabstate}\\
    g_1 = -\gamma_{A_1}^1 \gamma_{A_1}^2 \gamma_{A_2}^1\gamma_{A_3}^2\n
    g_2 = -\gamma_{A_1}^2 \gamma_{A_2}^1\gamma_{A_2}^2\gamma_{A_3}^1\n
    g_1 g_2 = -\gamma_{A_1}^1 \gamma_{A_2}^2 \gamma_{A_3}^1 \gamma_{A_3}^2\notag
\end{gather}
Here the subscript on the Majorana operators indicate the party they act on, while the superscript is a flavor index. For each partition $m|\overline{m}$ of the three parties, there are at least two terms in (\ref{eq:form_of_tripartite_stabstate}) that violate fermion parity across the partition. Hence, any biseparable product state resource $\sigma_m \otimes \sigma_{\overline{m}}$ will have a nonzero trace over only one of the non-identity terms. In particular, measurement outcomes using the resource state $\sigma_{A_1} \otimes \sigma_{A_2, A_3}$ can only depend on $s_1$, outcomes using the resource state $\sigma_{A_2} \otimes \sigma_{A_1,A_3}$ can only depend on $s_2$, and outcomes using $\sigma_{A_3} \otimes \sigma_{A_1, A_2}$ can only depend on $s_1 s_2$. None of these three resource states can distinguish both phases at the same time. 

Any biseparable resource state is a positive mixture of the above three types of product states; therefore, their probability distributions under a given POVM, $p(t_1, t_2, s_1, s_2) = \tr\big(M_{t_1 t_2} (\rho(s_1, s_2) \otimes \sigma)\big)$, must also be a positive mixture of probability distributions that use the biseparable product states as resources. In order to produce a delta distribution $p(t_1, t_2, s_1, s_2) = \delta_{t_1, s_1} \delta_{t_2,s_2}$, those individual distributions on the product states must also be the same delta distribution, which as we have seen is impossible. Therefore, biseparable resource states cannot be used to decode $s_1$ and $s_2$ from the state (\ref{eq:form_of_tripartite_stabstate}). A fermionic GME state that can be used to decode $s_1$ and $s_2$ is, similar to before, a copy of $\rho(+,+)$:
\begin{gather*}
    \sigma = \frac{1}{8}(I+g_1')(I+g_2') = \frac{1}{8}(I-\gamma_{A_1}^3 \gamma_{A_1}^4 \gamma_{A_2}^3\gamma_{A_3}^4)(I-\gamma_{A_1}^4 \gamma_{A_2}^3\gamma_{A_2}^4\gamma_{A_3}^3)\\
    M_{t_1, t_2} = \frac{1}{4}(I+t_1 g_1 g_1')(I+t_2 g_2 g_2')\\
    \tr\big(M_{t_1,t_2}(\rho(s_1,s_2) \otimes \sigma)\big) = \delta_{t_1,s_1} \delta_{t_2,s_2}
\end{gather*}

\section{Properties of the fermionic partial transpose} \label{app:fpt_identities}

In this section we will list some miscellaneous identities regarding the fermionic partial transpose $R_m$ and its complex conjugate $R_m^*$, which can be defined in terms of Majorana operators by sending all odd Majorana fermions $\gamma_{2j-1}$ to $i\gamma_{2j-1}$ instead of $-i\gamma_{2j-1}$, and sending all even Majorana fermions $\gamma_{2j}$ to $-i\gamma_{2j}$ instead of $i\gamma_{2j}$, for $j \in m$. 

\begin{description}
\item[Inverse] Firstly, by definition, the two types of fermionic partial transpose are inverses of each other, that is, $(A^{R_m})^{R_m^*} = (A^{R_m^*})^{R_m} = A$. 

\item[Self-Adjointness] The second identity is that the fermionic partial transpose and its complex conjugate are both individually self-adjoint, that is,
\begin{gather}
    \tr(A^{R_m} B) = \tr(A B^{R_m})\\
    \tr(A^{R_m^*} B) = \tr(A B^{R_m^*})
\end{gather}
which follows from the definition of $R_m$ in terms of Majorana fermions. 

\item[Hermitian conjugate] The fermionic partial transpose and its inverse exchange places under a Hermitian conjugation:
\begin{gather}
    \left(A^{R_m}\right)^\dagger = \left(A^\dagger\right)^{R_m^*}\\
    \left(A^{R_m^*}\right)^\dagger = \left(A^\dagger\right)^{R_m}
\end{gather}
This again follows from the definition of $R_m$ in terms of Majorana fermions. 

\item[Product rule] Proposition 2 of Ref.~\onlinecite{Shapourian2019} shows that the fermionic partial transpose on a product of states can be easily decomposed, provided all but one are separable along the partition:
\begin{gather}
    \left[(A_m \otimes A_{\overline{m}}) B (C_m \otimes C_{\overline{m}})\right]^{R_m} = (C_m^R \otimes A_{\overline{m}}) B^{R_m} (A_m^R \otimes C_{\overline{m}})\\
    \left[(A_m \otimes A_{\overline{m}}) B (C_m \otimes C_{\overline{m}})\right]^{R_{\overline{m}}} = (A_m \otimes C_{\overline{m}}^R) B^{R_{\overline{m}}} (C_m \otimes A_{\overline{m}}^R)
\end{gather}

\item[Positivity] Finally, we note that a full fermionic transpose $\rho^R$ preserves positivity if the original state preserves fermion parity and is positive semidefinite. This is because, when the fermion parity is preserved, the fermionic transpose is identical to the non-fermionic transpose under local unitaries (using the form (9) of the fermionic partial transpose of Ref.~\onlinecite{Shapourian2017}), which preserves positivity.
\end{description}

\section{Dual SDP to the fermionic GMN} \label{app:dual_sdp}
In this section we will derive the form of the dual problem to the fermionic GMN SDP (\ref{eq:dual_sdp_1}-\ref{eq:dual_sdp_2}).
As a first step, we rewrite the fermionic GMN SDP problem in its generic form 
\begin{gather*}
    \mathcal{N}^F = -\inf_{\vec{x} \in \mathbb{R}^n} \vec{c}^{\,T} \cdot \vec{x}
\end{gather*}
where $\vec{c}$ is a vector of parameters, and $\vec{x}$ is subject to the conditions
\begin{gather} \label{eq:f_condition_sdp}
    F(\vec{x}) = F_0 + \sum_i x_i F_i \succeq 0
\end{gather}
Taking a Hermitian complete orthogonal basis of the operators $\{\sigma_i\}$ with $\tr(\sigma_i \sigma_j) = \delta_{ij}, A = \sum_i \tr(A \sigma_i) \sigma_i$, and indexing the partitions $m$ by integers from $1$ to $M$,
the variable $\vec{x}$ and parameter $\vec{c}$ can be written as
\begin{gather*}
    \vec{x} = (\vec{w}_R, \vec{w}_I,\vec{p}^{\,1}_R, \vec{p}^{\,1}_I ,\vec{p}^{\,2}_R, \vec{p}^{\,2}_I ... \vec{p}^{\,M}_R,\vec{p}^{\,M}_I)\\
    \vec{c} = (\vec{\rho}, \vec{0} ... \vec{0})
\end{gather*}
with $\vec{w}_R = \re\tr(W\vec{\sigma}), \vec{w}_I = \im\tr(W\vec{\sigma})$ (such that $W = \sum_i (w_{Ri} + iw_{Ii})\sigma_i$), and $\vec{p}^{\,m}_{R} = \re \tr(P_m \vec{\sigma}), \vec{p}^{\,m}_{I} = \im \tr(P_m \vec{\sigma}), \vec{\rho} = \tr(\rho \vec{\sigma})$. Next, we look at the SDP condition matrix $F(\vec{x})$, which can account for all the fully fermion-decomposable conditions (\ref{eq:fermion_decomposable_condition_1}-\ref{eq:fermion_decomposable_condition_3}) if it takes the block diagonal form
\begin{align}
    F(\vec{x}) =& \left(\bigoplus_{m=1}^M P_m+P_m^\dagger \right) \oplus \left( \bigoplus_{m=1}^M\begin{pmatrix} I & 2(W-P_m)^{R_m^*} - I\\(2(W-P_m)^{R_m^*} - I)^\dagger & I \end{pmatrix}\right)\n
    =& \left(\bigoplus_{m=1}^M P_m+P_m^\dagger \right) \oplus \left( \bigoplus_{m=1}^M\begin{pmatrix} I & 2(W-P_m)^{R_m^*} - I\\2[(W-P_m)^\dagger]^{R_m} - I & I \end{pmatrix}\right) \label{eq:f_condition_sdp2}
\end{align}
We rewrite (\ref{eq:f_condition_sdp}) as 
\begin{gather}
    F(\vec{x}) = F_0 + \sum_i ( w_{Ri} F_{w_{Ri}} + w_{Ii} F_{w_{Ii}}) + \sum_{i,m} (p^m_{Ri} F_{p^m_{Ri}} + p^m_{Ii} F_{p^m_{Ii}}) \succeq 0.
\end{gather}
The matrices $F_0, F_{\vec{w}_R}, F_{\vec{w}_I}, F_{\vec{p}^m_R}, F_{\vec{p}^m_I}$ then have the block diagonal form 
\begin{gather*}
    F_0 = \left(\bigoplus_{m=1}^M 0 \right) \oplus \left(\bigoplus_{m=1}^M \begin{pmatrix} I & -I \\ -I & I \end{pmatrix} \right)\\
    F_{w_{Ri}} = \left(\bigoplus_{m=1}^M 0 \right) \oplus \left(\bigoplus_{m=1}^M \begin{pmatrix} 0 & 2\sigma_i^{R_m^*}\\2\sigma_i^{R_m} & 0 \end{pmatrix} \right)\\
    F_{w_{Ii}} = i\left(\bigoplus_{m=1}^M 0 \right) \oplus \left(\bigoplus_{m=1}^M \begin{pmatrix} 0 & 2\sigma_i^{R_m^*}\\-2\sigma_i^{R_m} & 0 \end{pmatrix} \right)\\
    F_{p^m_{Ri}} = \left(\bigoplus_{m'=1}^M 2\sigma_i \delta_{m m'} \right) \oplus \left(\bigoplus_{m'=1}^M \delta_{m m'}\begin{pmatrix} 0 & -2\sigma_i^{R_{m'}^*}\\-2\sigma_i^{R_{m'}} & 0 \end{pmatrix} \right)\\
    F_{p^m_{Ii}} = i\left(\bigoplus_{m'=1}^M 0 \right) \oplus \left(\bigoplus_{m'=1}^M \delta_{m m'}\begin{pmatrix} 0 & -2\sigma_i^{R_{m'}^*}\\2\sigma_i^{R_{m'}} & 0 \end{pmatrix} \right)\\
\end{gather*}

The dual SDP problem is 
\begin{gather}
    \mathcal{N}^F = \inf_Z \tr(F_0 Z)
\end{gather}
subject to
\begin{gather*}
    Z \succeq 0\\
    \tr(F_i Z) = c_i \quad\forall i
\end{gather*}
or in our terms
\begin{gather*}
    \tr(F_{w_{Ri}} Z) = \rho_i\\
    \tr(F_{w_{Ii}} Z) = \tr(F_{P^m_{Ri}} Z) = \tr(F_{P^m_{Ii}} Z) = 0.
\end{gather*}
The relevant terms of $Z$ lie on the same block diagonal
\begin{gather*}
    Z = \left(\bigoplus_{m=1}^M Z_m \right) \oplus \left(\bigoplus_{m=1}^M \begin{pmatrix} V_m & Y_m\\Y_m^\dagger & X_m \end{pmatrix} \right)
\end{gather*}
where all $Z_m$ and $\begin{pmatrix} V_m & Y_m\\Y_m^\dagger & X_m \end{pmatrix}$ must each be positive semidefinite. The condition $\tr(F_{w_{Ri}} Z) = \rho_i$ becomes
\begin{gather*}
    \rho_i = \sum_m \tr(2 \sigma_i^{R_m^*} Y_m^\dagger + 2\sigma_i^{R_m} Y_m)
\end{gather*}
which can be cleaned up using the self-adjoint nature of $R_m$ and $R_m^*$ to
\begin{align*}
    \rho_i &= \sum_m \tr(2 \sigma_i (Y_m^\dagger)^{R_m^*} + 2\sigma_i^{R_m} Y_m)\\
    &= \sum_m \tr\big( 2\sigma_i [(Y_m^{R_m})^\dagger + Y_m^{R_m}]\big)
\end{align*}
The condition $\tr(F_{P_{Ii}^m} Z) = 0$ becomes
\begin{align*}
    i\,\tr(2 \sigma_i^{R_m^*} Y_m^\dagger - 2\sigma_i^{R_m} Y_m) &= 0\\
    \tr\big( 2\sigma_i [(Y_m^{R_m})^\dagger - Y_m^{R_m}]\big) &= 0\\
    (Y_m^{R_m})^\dagger - Y_m^{R_m} &= 0,
\end{align*}
that is, $Y_m^{R_m}$ must be Hermitian. The condition $\tr(F_{w_{Ii}} Z) = 0$ also reduces to this, but summed over $m$. Finally, the condition $\tr(F_{P_{Ri}^m} Z) = 0$ becomes
\begin{align*}
    2\tr(\sigma_i Z_m) &= \tr(2 \sigma_i^{R_m^*} Y_m^\dagger + 2\sigma_i^{R_m} Y_m)\\
    &= \tr\big( 2\sigma_i [(Y_m^{R_m})^\dagger + Y_m^{R_m}]\big)\\
    Z_m &= (Y_m^{R_m})^\dagger + Y_m^{R_m}
\end{align*}
Which, since $Y_m^{R_m}$ must be Hermitian, becomes
\begin{gather}
    Z_m = 2Y_m^{R_m}.
\end{gather}
Therefore $\rho_i = \sum_m 2\tr(\sigma_i Z_m)$, so
\begin{align*}
    \rho &= \sum_i \rho_i \sigma_i \\
    &= \sum_{i,m} 2\tr(\sigma_i Z_m) \sigma_i\\
    &= \sum_m 2Z_m
\end{align*}
As each $Z_m$ must be positive semidefinite, it can be written as $\half p_m \rho_m$, where $p_m > 0$ and $\rho_m$ is a positive semidefinite operator with $\tr(\rho_m) = 1$. Therefore, after rescaling $X_m, V_m, Y_m$ by $p_m$, the dual SDP problem becomes
\begin{gather}
    \tr(F_0 Z) = \sum_m p_m \tr(V_m + X_m - Y_m - Y_m^\dagger)\\
    \mathcal{N}^F = \inf_{p_m, \rho_m} \sum_m p_m \inf_{V_m, X_m, Y_m} \tr(V_m+X_m - Y_m - Y_m^\dagger)\label{eq:dual_sdp}
\end{gather}
subject to the conditions
\begin{gather*}
    \rho = \sum_m p_m \rho_m, \qquad \rho_m = 4Y_m^{R_m} \succeq 0, \qquad \begin{pmatrix} V_m & Y_m\\Y_m^\dagger & X_m \end{pmatrix} \succeq 0
\end{gather*}
We can rewrite the expression in the 2nd infimum of (\ref{eq:dual_sdp}) as $\mathcal{N}^F_m(\rho_m)$, such that
\begin{gather}\label{eq:dual_sdp_function_of_NF}
    \mathcal{N}^F(\rho) = \inf\bigg\{ \sum_m p_m \mathcal{N}^F_m(\rho_m)\,\bigg|\, \rho_m \succeq 0 \,\, \forall \, m, \text{ and } \rho = \sum_m p_m \rho_m\bigg\}
\end{gather}
Using $Y_m = \quarter \rho_m^{R_m^*}$,
\begin{align*}
    \mathcal{N}_m^F(\rho_m) &= \inf\bigg\{ \tr(V_m+X_m)-\tr\big(\quarter \rho_m^{R_m^*} + \quarter \rho_m^{R_m}\big)\,\,\bigg| \,\begin{pmatrix} V_m & \quarter \rho_m^{R_m^*}\\\quarter \rho_m^{R_m} & X_m \end{pmatrix} \succeq 0 \bigg\}
\end{align*}
As the fermionic partial transpose (and its conjugate) are trace preserving, this becomes
\begin{align*}
    \mathcal{N}_m^F(\rho_m) &= \inf \bigg\{ \tr(V_m+X_m)-\half \,\,\bigg|\, \begin{pmatrix} V_m & \quarter \rho_m^{R_m^*}\\\quarter \rho_m^{R_m} & X_m \end{pmatrix} \succeq 0 \bigg\}.
\end{align*}
Consider the singular value decomposition $Y_m = \quarter \rho_m^{R_m^*} =  \sum_i y_i \vec{a}_i \vec{b}_i^\dagger$, where $\vec{a}_i, \vec{b}_i$ are two orthonormal bases of the Hilbert space. We have
\begin{align*}
    \begin{pmatrix} \vec{a}_i^\dagger & -\vec{b}_i^\dagger \end{pmatrix} \begin{pmatrix} V_m & Y_m\\Y_m^\dagger & X_m \end{pmatrix} \begin{pmatrix} \vec{a}_i \\ -\vec{b}_i \end{pmatrix} &= \vec{a}_i^{\dagger} V_m \vec{a}_i + \vec{b}_i^\dagger X_m \vec{b}_i - \vec{b}_i^\dagger Y_m^\dagger a_i - \vec{a}_i^\dagger Y_m \vec{b}_i\\
    &=\vec{a}_i^{\dagger} V_m \vec{a}_i + \vec{b}_i^\dagger X_m \vec{b}_i - 2y_i \geq 0
\end{align*}
Summing over all $i$ therefore gives 
\begin{gather}
    \tr(V_m+X_m) \geq 2\tr(|Y_m|) = \half \tr(|\rho_m^{R_m}|)\n
    \mathcal{N}_m^F(\rho_m) \geq \half \tr(|\rho_m^{R_m}|)-\half \label{eq:fermion_bipartite_inequality}
\end{gather}
and these inequalities can be saturated, by setting $V_m = X_m = |Y_m|$, so the inequality (\ref{eq:fermion_bipartite_inequality}) is actually an equality, and
\begin{gather}\label{eq:dual_sdp_tracenorm}
    \mathcal{N}_m^F(\rho_m) = \half \tr(|\rho_m^{R_m}|)-\half
\end{gather}
We can see that (\ref{eq:dual_sdp_tracenorm}) is simply the bipartite negativity of $\rho_m$ across the partition $m|\overline{m}$. Combining with (\ref{eq:dual_sdp_function_of_NF}) gives the full expression of the fermionic GMN in the dual SDP as
\begin{gather}\label{eq:full_dual_sdp_tracenorm}
    \mathcal{N}^F = -\half + \half\inf\left\{ \sum_m p_m \tr(|\rho_m^{R_m}|)\,\bigg|\, \rho_m \succeq 0 \,\, \forall \, m, \text{ and } \rho = \sum_m p_m \rho_m\right\}
\end{gather}

\section{Entanglement of fermionic mixed stabilizer states} \label{app:stabilizer_fgmn}

Here we prove that the fermionic GMN of any RDM of a pure stabilizer state is equal, at all parties, to the minimum bipartite negativity across any nontrivial partition $m|\overline{m}$ of the parties (similar to pure states). Such a state would take the form
\begin{gather}
    \rho = 2^{-(N-S)} \prod_{s=1}^S \half(I+g_s)
\end{gather}
where the $S$ stabilizer generators $g_s$ are commuting Hermitian products of some even number $2\kappa_s$ of Majorana fermions: $g_s = \pm i^{\kappa_s} \gamma_{a_s^1} \gamma_{a_s^2}...\gamma_{a_s^{2 \kappa_s}}$.
To prove our claim, we take the set of $S$ stabilizers $g_s$ and construct a complete orthogonal family of projectors $\Pi_{\vec{u}}$ indexed by $\vec{u} \in \mathbb{Z}_2^{\times S}$:
\begin{gather}
    \Pi_{\vec{u}} = \prod_{s=1}^S \Pi_{s,u_s}\\
    \Pi_{s,u_s} = \half (I+(-1)^{u_s}g_s)
\end{gather}
noting that $\rho = 2^{-(N-S)} \Pi_{\vec{0}}$, $\tr(\Pi_{\vec{u}} \Pi_{\vec{v}}) = 2^{N-S} \delta_{\vec{u} \vec{v}}$, and $I = \sum_{\vec{u}} \Pi_{\vec{u}}$.
We also define the depolarizing channel around these projectors
\begin{gather}
    S[A] = 2^{-(N-S)}\sum_{\vec{u}} \tr(\Pi_{\vec{u}} A) \Pi_{\vec{u}} \equiv \sum_{\vec{u}} A_{\vec{u}} \Pi_{\vec{u}}.
\end{gather}
This channel preserves the trace with $\rho$, i.e.
\begin{align*}
    \tr(\rho S[W]) &= 2^{-(N-S)}\sum_{\vec{u}} W_{\vec{u}} \tr(\Pi_{\vec{0}} \Pi_{\vec{u}})\\
    &= W_{\vec{0}}\\
    &= \tr(\rho W)
\end{align*}
Our first step is to prove that if $W \in \mathcal{W}^F$, then so is $S[W]$. We have, for every partition $m|\overline{m}$, 
\begin{align}
    S[W] &= S[P_m] + S[Q_m^{R_m}]\n
    &= S[P_m] + (S[Q_m^{R_m}]^{R_m^*})^{R_m},
\end{align}
hence we will prove $S[P_m] + S[P_m]^\dagger \geq 0$ and $||2S[Q_m^{R_m}]^{R_m^*} - I|| \leq 1$. The former comes from the observation that $S$ is linear, $S[P_m]^\dagger = S[P_m^\dagger]$, and $\tr[\Pi_{\vec{u}} (P_m+P_m^\dagger)] \geq 0$ for all $\vec{u}$, so $S[P_m] + S[P_m]^\dagger$ is just a positive mixture of projectors. For the latter proof, we expand as
\begin{align}
    S[Q_m^{R_m}]^{R_m^8} &= 2^{-(N-S)} \sum_{\vec{u}} \tr(\Pi_{\vec{u}} Q_m^{R_m}) \Pi_{\vec{u}}^{R_m^*}\n
    &= 2^{-(N-S)} \sum_{\vec{u}} \tr(\Pi_{\vec{u}}^{R_m} Q_m) \Pi_{\vec{u}}^{R_m^*}
\end{align}
Here we can use the specific nature of the $\Pi_{\vec{u}}$ projectors. The partial fermionic transpose over the set $m$ sends each odd Majorana fermion $\gamma_{2a-1}$ to $-i\gamma_{2a-1}$, and each even Majorana fermion $\gamma_{2a}$ to $-i\gamma_{2a}$, if $a \in m$. Therefore, by expanding the product in $\Pi_{\vec{u}}$,
\begin{gather}
    \Pi_{\vec{u}} = 2^{-S} \sum_{\vec{x} \in \{0,1\}^S} (-1)^{\vec{u} \cdot \vec{x}} g_1^{x_1}...g_S^{x_S}\\
    \Pi_{\vec{u}}^{R_m} = 2^{-S} \sum_{\vec{x} \in \{0,1\}^S} (-1)^{\vec{u} \cdot \vec{x}} i^{\kappa^m(\vec{x})} g_1^{x_1}...g_S^{x_S}
\end{gather}
where $\kappa^m(\vec{x})$ is the total number of even Majorana fermions of the product $g_1^{x_1}...g_S^{x_S}$ contained in subregion $m$, minus the total number of odd Majorana fermions. Note that this is not a sum of the individual Majorana fermion count disparities with respect to each generator, because fermions from some generators might cancel out with those from another. Instead, this can be written as $\vec{\kappa}^m \cdot \vec{x} - 2\theta^m(\vec{x})$, where $(\vec{\kappa}^m)_s = \kappa^m(\vec{e}_s)$ and $\theta^m(\vec{x})$ is the cancellation parameter, accounting for how many times that Majorana fermions in subregion $m$ cancel out in the product $g_1^{x_1}...g_S^{x_S}$. Therefore, we have
\begin{align*}
    \Pi_{\vec{u}}^{R_m} &= 2^{-S} \sum_{\vec{x} \in \{0,1\}^S} (-1)^{\vec{u} \cdot \vec{x} + \theta^m(\vec{x})} i^{\vec{\kappa}^m\cdot \vec{x}} g_1^{x_1}...g_S^{x_S}\\
    &= 2^{-S} \sum_{\vec{x} \in \{0,1\}^S} (-1)^{\vec{u} \cdot \vec{x} + \theta^m(\vec{x})} i^{\vec{\kappa}^m\cdot \vec{x}} \prod_{s=1}^S (\Pi_{s,0}+(-1)^{x_s} \Pi_{s,1})\\
    &= 2^{-S} \sum_{\vec{x}} (-1)^{\vec{u} \cdot \vec{x} + \theta^m(\vec{x})} i^{\vec{\kappa}^m\cdot \vec{x}} \sum_{\vec{v}} (-1)^{\vec{x}\cdot \vec{v}} \Pi_{\vec{v}}
\end{align*}
Which means that in all cases, the projector $\Pi_{\vec{u}}$ is sent to a linear combination of other projectors 
\begin{gather}
    \Pi_{\vec{u}}^{R_m} = \sum_{\vec{v}}\mu^m_{\vec{u} \vec{v}} \Pi_{\vec{v}}
\end{gather} 
under fermionic partial transpose on set $m$, with
\begin{gather}
    \mu^m_{\vec{u}{\vec{v}}} = 2^{-S} \sum_{\vec{x}} (-1)^{\vec{x}\cdot(\vec{u}+\vec{v}) + \theta^m(\vec{x})} i^{\vec{\kappa}^m\cdot \vec{x}} \label{eq:stabilizer_fpt_coefficient_general_form}
\end{gather}
In the special case that all stabilizer generators locally commute,  $\theta^m(\vec{x})$ is always even, so we can drop its phase factor out of (\ref{eq:stabilizer_fpt_coefficient_general_form}), simplifying it to
\begin{gather}
    \mu^m_{\vec{u}{\vec{v}}} = \prod_{s=1}^S \frac{1+(-1)^{u_s+v_s} i^{\kappa^m_s}}{2} \label{eq:stabilizer_fpt_coefficient_precise_form}
\end{gather}
Now we will use properties of the fermionic partial transpose to infer properties of the coefficients $\mu^m_{\vec{u} \vec{v}}$. Firstly, from the definition of the conjugate transpose $R_m^*$ in terms of the Majorana operators, we have 
\begin{gather}
    \Pi_{\vec{u}}^{R_m^*} = \sum_{\vec{v}}(\mu^m_{\vec{u} \vec{v}})^* \Pi_{\vec{v}}
\end{gather}
Because the fermionic partial transpose preserves trace,
\begin{gather}
    \tr(\Pi_{\vec{u}}) = \tr(\Pi_{\vec{u}}^{R_m}) = \sum_{\vec{v}}\mu^m_{\vec{u} \vec{v}} \tr(\Pi_{\vec{v}})\n
    \sum_{\vec{v}}\mu^m_{\vec{u} \vec{v}} = 1
\end{gather}
Because the conjugate transpose cancels the original fermionic partial transpose, we have
\begin{gather}
    \sum_{\vec{v}} (\mu^m_{\vec{u}\vec{v}})^* \mu^m_{\vec{v} \vec{w}} = \delta_{\vec{u} \vec{w}}
\end{gather}
Because the fermionic partial transpose is self-adjoint,
\begin{align}
    \mu^m_{\vec{u}\vec{v}} &= 2^{-(N-S)} \tr(\Pi_{\vec{u}}^{R_m} \Pi_{\vec{v}})\n
    &= 2^{-(N-S)} \tr(\Pi_{\vec{u}} \Pi_{\vec{v}}^{R_m})\n
    &= \mu^m_{\vec{v}\vec{u}}
\end{align}
Combining these identities, we have
\begin{align}
    S[Q_m^{R_m}]^{R_m^*} &= 2^{-(N-S)} \sum_{\vec{u}} \tr(\Pi_{\vec{u}}^{R_m} Q_m) \Pi_{\vec{u}}^{R_m^*}\n
    &= \sum_{\vec{u} \vec{v}\vec{w}} \mu^m_{\vec{u}\vec{v}} (\mu^m_{\vec{u}\vec{w}})^* Q_{m \vec{v}} \Pi_{\vec{w}}\n
    &= \sum_{ \vec{v}\vec{w}} Q_{m \vec{v}} \Pi_{\vec{w}} \sum_{\vec{u}}(\mu^m_{\vec{w}\vec{u}})^* \mu^m_{\vec{u}\vec{v}} \n
    &=  \sum_{ \vec{u}} Q_{m \vec{u}} \Pi_{\vec{u}}\\
    2S[Q_m^{R_m}]^{R_m^*}-I &= \sum_{ \vec{u}} (2Q_{m \vec{u}}-1) \Pi_{\vec{u}}
\end{align}
As a sum of orthogonal projectors, the operator norm of the final expression is bounded by its maximum coefficient,
\begin{align}
    \max_{\vec{u}} |2Q_{m \vec{u}}-1| &= \max_{\vec{u}} |2^{-(N-S)}\tr[(2Q_m-I) \Pi_{\vec{u}}]| 
\end{align}
Decomposing $\Pi_{\vec{u}}$ as a sum of $2^{N-S}$ basis elements $|i\rangle \langle i|$, and using the original operator norm bound on $2Q_m-I$, therefore gives $|2Q_{m \vec{u}}-1|\leq 1$, yielding the operator norm bound $||2S[Q_m^{R_m}]^{R_m^*}-I|| \leq 1$, so all the requirements for $S[W]$ to be a member of $\mathcal{W}^F$ are fulfilled.

Because $S[W] \in \mathcal{W}^F$ and has the same overlap with $\rho$ as the original witness $W$, we can restrict our search over $S[W]$ for all $W \in \mathcal{W}^F$. Because
\begin{align}
    S[W] &= \sum_{\vec{u}} W_{\vec{u}} \Pi_{\vec{u}}\\
    S[P_m] + S[Q_m^{R_m}]&= \sum_{\vec{u}} \left(P_{m \vec{u}} + \sum_{\vec{v}} \mu^m_{\vec{u}\vec{v}} Q_{m \vec{v}}\right)\Pi_{\vec{u}} \qquad \forall m
\end{align}
our SDP problem can be decomposed according to the projectors and the coefficients of $W, P_m, Q_m$ on these projectors:
\begin{gather}
    \mathcal{N}^F = -\inf_{W_{\vec{u}}} \re W_{\vec{0}}
\end{gather}
subject to the conditions
\begin{gather}
    W_{\vec{u}} = P_{m \vec{u}} + \sum_{\vec{v}} \mu^m_{\vec{u}\vec{v}} Q_{m \vec{v}}\label{eq:w_sum_condition}\\
    \re P_{m \vec{u}} \geq 0\\
    |2Q_{m \vec{u}}-1| \leq 1
\end{gather}
We will try to optimize over one condition of (\ref{eq:w_sum_condition}) first, the $\vec{u} = 0$ condition over a single partition $m$ - that is, find the infimum of 
\begin{align}
    \re W_{\vec{0}} &= \re P_{m \vec{0}} + \re \sum_{\vec{v}}\mu^m_{\vec{0}\vec{v}} Q_{m \vec{v}}\n
    &= \re P_{m \vec{0}} + \half + \half\sum_{\vec{v}}\re \mu^m_{\vec{0}\vec{v}} (2Q_{m \vec{v}}-1)
\end{align}
which is clearly minimized, given the conditions on $P_{m \vec{u}}$ and $2Q_{m \vec{u}}-1$, when $P_{m \vec{0}}$ is any value with zero real component and $2Q_{m \vec{v}}-1 = -(\mu^m_{\vec{0}\vec{v}})^*/|\mu^m_{\vec{0}\vec{v}}|$ (if $\mu^m_{\vec{0}\vec{v}}=0$ we will just set $2Q_{m \vec{v}}-1 = 0$ as well), giving
\begin{align}
    \re W_{\vec{0}} &= \half - \half\sum_{\vec{v}}|\mu^m_{\vec{0}\vec{v}}|
\end{align}
In order to make (\ref{eq:w_sum_condition}) consistent for all partitions $m$, we need to rescale some $2Q_{m \vec{v}}-1$ values. Taking $T_m = \sum_{\vec{v}}|\mu^m_{\vec{0}\vec{v}}|$, we achieve consistency with minimum possible $\re W_{\vec{0}}$ at 
\begin{gather}
    2Q_{m \vec{v}}-1 \rightarrow (2Q_{m \vec{v}}-1)\frac{\min_{m'} T_{m'}}{T_m} \label{eq:qm_coefficient_rescaling_condition}
\end{gather}
That is, ignoring the $\vec{u} \neq \vec{0}$ conditions in (\ref{eq:w_sum_condition}), we have
\begin{gather}
    \mathcal{N}^F = \half \min_{m} \sum_{\vec{v}}|\mu^m_{\vec{0}\vec{v}}| - \half \label{eq:min_negativity_from_transpose_coefficients}
\end{gather}
Noting that
\begin{gather}
    \rho^{R_m} = 2^{-(N-S)}\Pi_{\vec{0}}^{R_m} = 2^{-(N-S)}\sum_{\vec{v} }\mu^m_{\vec{0}\vec{v}} \Pi_{\vec{v}}\n
    \tr(|\rho^{R_m}|) = \sum_{\vec{v} }|\mu^m_{\vec{0}\vec{v}}|
\end{gather}
we see that (\ref{eq:min_negativity_from_transpose_coefficients}) is precisely the minimum bipartite fermionic negativity across all partitions. 

It remains to show that the $\vec{u} \neq \vec{0}$ conditions in (\ref{eq:w_sum_condition}) can be satisfied without modifying the values of $W_{\vec{0}}$, $P_{m \vec{0}}$ and $Q_{m \vec{u}}$. Substituting $2Q_{m \vec{v}}-1 = -q_m(\mu^m_{\vec{0}\vec{v}})^*/|\mu^m_{\vec{0}\vec{v}}|$ where $q_m$ is the rescaling factor from (\ref{eq:qm_coefficient_rescaling_condition}), our conditions become
\begin{align}
    W_{\vec{u}} &= P_{m\vec{u}} + \half - \half q_m \sum_{\vec{v}}  \frac{(\mu^m_{\vec{0}\vec{v}})^*\mu^m_{\vec{u}\vec{v}}}{|\mu^m_{\vec{0}\vec{v}}|}\label{eq:w_sum_condition_2}
\end{align}
which can be satisfied by choosing
\begin{gather}
    \re W_{\vec{u}} > \max_m \re \left(\half - \half q_m \sum_{\vec{v}}  \frac{(\mu^m_{\vec{0}\vec{v}})^*\mu^m_{\vec{u}\vec{v}}}{|\mu^m_{\vec{0}\vec{v}}|}\right)\\
    P_{m \vec{u}} = W_{\vec{u}} - \half + \half q_m \sum_{\vec{v}}  \frac{(\mu^m_{\vec{0}\vec{v}})^*\mu^m_{\vec{u}\vec{v}}}{|\mu^m_{\vec{0}\vec{v}}|}
\end{gather}
as the only relevant restriction on $P_{m \vec{u}}$, that its real part is nonnegative, can be easily satisfied by choosing a large enough $W_{\vec{u}}$, with no penalty to the negativity.

If all stabilizers locally commute, such as in the Kitaev model of Section~\ref{sec:kitaev_ground_state}, we can apply (\ref{eq:stabilizer_fpt_coefficient_precise_form}) to (\ref{eq:min_negativity_from_transpose_coefficients}) to get a simple expression for $\mathcal{N}^F$ in terms of the stabilizers, in terms of the set $S^o_m$ of stabilizers $s$ where $\kappa^m_s$ is odd:
\begin{align}
    \mathcal{N}^F &= \half \min_{m} \sum_{\vec{v}}|\mu^m_{\vec{0}\vec{v}}| - \half\n
    &= \half \min_m \sum_{\vec{v}} \prod_{s=1}^S  \left|\frac{1+i^{\kappa^m_s}}{2} \delta_{v_{s} 0} + \frac{1-i^{\kappa^m_s}}{2} \delta_{v_{s} 1}\right| - \half\n 
    &= \half \min_m \prod_{s \in S_m^o}\ \left(\left|\frac{1+i}{2}\right| +\left|\frac{1-i}{2} \right|\right)-\half\n 
    &= \half \min_m \sqrt{2}^{|S_m^o|} - \half
\end{align}
Effectively, our entanglement depends on the minimum number of stabilizers ``cut" by any partition $m$, where a stabilizer is ``cut" if it violates fermion parity locally on $m$.

\section{Biseparable decomposition of simple mixed states}\label{app:bisep_forms}
Here we will list the biseparable decompositions corresponding to the known sudden death points in Table~\ref{tab:white_noise_sudden_deaths}. We start with the simple 2-3 qubit entangled states
\begin{gather}
    \ket{\Phi^\pm} = \frac{1}{\sqrt{2}} \big(|00\rangle \pm |11\rangle\big)\\
    \ket{\Psi^\pm} = \frac{1}{\sqrt{2}} \big(|10\rangle + |01\rangle\big)\\
    \ket{G^1} = \frac{1}{2} \big(\ket{001} + \ket{010} + \ket{100} + \ket{111}\big)\\
    \ket{G^2} = \frac{1}{2} \big(\ket{000} + \ket{011} + \ket{101} + \ket{110}\big)
\end{gather}
and their density matrix forms $\Phi^\pm, \Psi^\pm, G^1, G^2$, etc., as well as $0_A = \ket{0}\bra{0}_A, 1_A = \ket{1}\bra{1}_A$. Note that all these states preserve fermion parity.
We have the following biseparable decomposition for the 3-qubit $X$-basis GHZ with $\frac{4}{5}$ mixing:
\begin{align}
    \frac{1}{5}\ket{GHZ_3}\bra{GHZ_3} + \frac{4}{5}(I/8) = \frac{1}{10}\big( &\Phi_{AB}^+\otimes 0_C + \Psi_{AB}^+ \otimes 1_C\n
    +& \Phi_{AC}^+\otimes 0_B  + \Psi_{AC}^- \otimes 1_B \n
    +& 0_A \otimes \Phi_{BC}^+ + 1_A \otimes \Psi_{BC}^+ \big) + (diag.)
\end{align}
where $(diag.)$ indicates miscellaneous density matrix components which are diagonal in the computational basis, and therefore inherently separable - in this case the diagonal terms with a different parity from the $\ket{GHZ_3}$ state. The 3-qubit W state with $\frac{8}{11}$ mixing decomposes to
\begin{gather}
    \frac{3}{11}\ket{W_3}\bra{W_3} + \frac{8}{11}(I/8) = \frac{2}{11}\left( \Psi_{AB}^+ \otimes 0_C + \Psi_{AC}^+ \otimes 0_B + 0_A \otimes \Psi_{BC}^+ \right) + (diag.)
\end{gather}
The 4-qubit $X$-basis GHZ with $\frac{8}{11}$ mixing decomposes to
\begin{align}
    \frac{1}{5}\ket{GHZ_4}\bra{GHZ_4} + \frac{4}{5}(I/16) = \frac{1}{20}\bigg(&\sum_{s=\pm} \Phi^s_{AB} \otimes \Phi^s_{CD} + \Psi^s_{AB} \otimes \Psi^s_{CD}\n
    &+\sum_{cyc.}(G^1_{ABC}\otimes 1_D + G^2_{ABC}\otimes 0_D\bigg) + (diag.)
\end{align}
where $cyc.$ indicates a cyclic sum over the parties $A,B,C,D$, changing the signs of the terms in $G^1$ to keep all its terms positive after reordering the sites. Finally, for the 4-qubit $Z$-basis GHZ state with $\frac{8}{11}$ mixing, we define
\begin{gather}
    \ket{\Phi^z} = \frac{1}{\sqrt{2}}\big(\ket{00} + z\ket{11}\big)
\end{gather}
giving the biseparable decomposition
\begin{align}
    \frac{3}{11}\ket{Z-GHZ_4}&\bra{Z-GHZ_4} + \frac{8}{11}(I/16) = \n
    &\frac{1}{22} \sum_{s=1}^4 \Phi^{i^s}_{AB} \otimes \Phi^{i^{-s}}_{CD} + \Phi^{i^s}_{AC} \otimes \Phi^{-i^{-s}}_{BD} + \Phi^{i^s}_{AD} \otimes \Phi^{i^{-s}}_{BC} + (diag.)
\end{align}

\end{appendices}

\end{document}